\documentclass[12pt]{article}
\usepackage{graphicx}
\newcommand{\beq}{\begin{equation}}
\newcommand{\eeq}{\end{equation}}
\newcommand{\bea}{\begin{eqnarray}}
\newcommand{\eea}{\end{eqnarray}}

\def\fun#1#2{\lower3.6pt\vbox{\baselineskip0pt\lineskip.9pt
  \ialign{$\mathsurround=0pt#1\hfil##\hfil$\crcr#2\crcr\sim\crcr}}}
\begin{document}
\begin{titlepage}
\begin{flushleft}
%       \hfill                      {\tt hep-th/03010***}\\
       \hfill                       FIT HE - 04-02 \\
\end{flushleft}
\vspace*{3mm}
\begin{center}
{\bf\LARGE Radius stabilization and brane running in RS1 model\\ }
%{\bf\LARGE
% for brane-world \\ }
\vspace*{5mm}

\bigskip

{\large Iver Brevik \footnote{\tt iver.h.brevik@mtf.ntnu.no}\\}
\vspace{2mm}
{
%\large
Department of Energy and Process Engineering, Norwegian University of Science and Technology,
N-7491 Trondheim, Norway\\}
\vspace*{5mm}

{\large Kazuo Ghoroku\footnote{\tt gouroku@dontaku.fit.ac.jp}\\ }
\vspace*{2mm}
{
%\large
${}^2$Fukuoka Institute of Technology, Wajiro, Higashi-ku}\\
{
%\large
Fukuoka 811-0295, Japan\\}
\vspace*{5mm}

{\large Masanobu Yahiro \footnote{\tt yahiro@sci.u-ryukyu.ac.jp} \\}
\vspace{2mm}
{
%\large
${}^4$Department of Physics and Earth Sciences, University of the Ryukyus,
Nishihara-chou, Okinawa 903-0213, Japan \\}

\vspace*{10mm}

\end{center}

\begin{abstract}
We study the effective potential of a scalar field based on the 5D
gauged supergravity for the RS1 brane model in terms of the
brane running method. The scalar 
couples to the brane such that the BPS conditions are satisfied
for the bulk configuration. The resulting effective potential
implies that the interbrane distance is undetermined in this case,
and we need a small BPS breaking term on the brane to stabilize
the interbrane distance at a finite length. We also discuss the
relationship to the Goldberger-Wise model.

\end{abstract}
\end{titlepage}

\section{Introduction}
It is quite probable that our 4d world is formed from a three
brane embedded in a higher dimensional space, probably according
to a ten-dimensional superstring theory. As such a model, we can
consider a thin three-brane (Randall-Sundrum brane) embedded in
AdS$_5$ space ~\cite{RS1,RS2}. In this case  the coordinate
transverse to the brane is considered as the energy scale of the
field theory on the boundary, in the sense of the AdS/CFT
correspondence \cite{M1,GKP1,W1,Poly1}.

\vspace{.2cm} In the original Randall-Sundrum (RS) model, the
brane action is expressed in terms of a tension parameter only. It
has recently been pointed out that many higher derivative terms
appear in the brane action when we make a change of the brane
position without changing the solutions for the background and the
Kaluza-Klein (KK) modes \cite{LMS,LR,Re,U,BGY}. This procedure is
called  brane running, and is considered to be an approximate
method of obtaining renormalization group flows for the field
theory on the brane. This implies that the parameters of the
brane-action on a different position are related through this
flow. Inspired by this idea, we examine the running behavior of
the brane action for a scalar field and derive the effective
action to 
%%%%%%%%%%%% Modified 1
%find a quantum corrected background configuration 
examine the stability of the two brane system via the brane running method.
%%%%%%%%%%%%%%%%%
We should notice that the effective potential, given by Eq.(\ref{effaction4}), 
is here
estimated from gravity side classical potential by pulling the hidden brane
to the position of the visible one~\cite{LMS}.
This point is different from many other
approaches to explain the radius stabilization.
%%%%%%%%%%%%%%%%

\vspace{.2cm} We consider a scalar field of a 5d gauged
supergravity model, and  restrict ourselves to the case of one
scalar field  coupled to the brane in a form consistent with the
Bogomolnyi-Plasad-Sommerfield
(BPS) conditions. The flow equations for the brane action of the
scalar are solved in a nonlinear form with self-interactions, in
order to find a non-trivial solution. It is shown for the RS1 (the
two brane model) that the effective action implies  the interbrane
distance to be arbitrary for a BPS solution. When the BPS
condition is slightly broken through the brane action, the
distance between the two branes is stabilized
%%%%%%%%%%%%%%%%%%%%%%%%%%%%
in spite of the particular form of the potential used here.
%%%%%%%%%%%%%%%%%%%%%%%%%%% 
The model of
Goldberger-Wise for the effective potential is also discussed in
our context.

\vspace{.2cm}
In Section 2, the model used here is set, and the flow equation for the
brane is derived and solved in  Section 3.
In Section 4, a possible stability of the braneworld solutions
is examined.
Concluding remarks are given in the final section.

\section{Setting of RS1 brane-world}

As  bulk action, consider a model of 5d gauged supergravity in the
Einstein frame as\footnote{ Here we take the following definition,
$R_{\nu\lambda\sigma}^{\mu}
=\partial_{\lambda}\Gamma_{\nu\sigma}^{\mu}-\cdots$,
$R_{\nu\sigma}=R_{\nu\mu\sigma}^{\mu}$ and
$\eta_{AB}=$diag$(-1,1,1,1,1)$. Five dimensional suffices are
denoted by capital Latin and four dimensional ones by  Greek
letters.
} %given in \cite{GPPZ},
\beq
   S_{\rm g}=\int d^4\!xdy\sqrt{-g}
   \left\{{1\over 2\kappa^2}R
    -{1\over 2}\sum_I(\partial\phi_I)^2-V(\phi)\right\}
          +{2\over 2\kappa^2}\int d^4x\sqrt{-g}K \ ,
                                                     \label{ac1g}
\eeq where $K$ is  the extrinsic curvature on the boundary. The
five-dimensional gravitational constant $\kappa^2$ is taken to be
$\kappa^2=2$ for simplicity. The potential $V$ is written in terms
of  a superpotential $W(\phi_I)$ as \beq
 V={v^2\over 8}\sum_I\left({\partial W\over \partial \phi_I}\right)^2
   -{v^2\over 3}W^2 . \label{superPot}
\eeq The gauge coupling parameter $v$ is fixed from the AdS$_5$
vacuum \cite{FGPW} in which $\phi_I=0$, and is given as $v=-2$ by
using the radius of AdS as a unit length. As for the brane action,
it is given such that it satisfies the BPS conditions on the
boundary \cite{G1}, \beq
    S_{\rm b} = -{v}\int d^4x dy\sqrt{-g}W(\phi_I)
\left(\delta(y-y_h)-\delta(y-y_v)\right). \label{baction}
\eeq
%\vspace{.3cm}
The background solutions are obtained under the following ansatz
for the metric, \beq
 ds^2= A^2(y)\eta_{\mu\nu}dx^{\mu}dx^{\nu}
           +dy^2  \, \label{metrica},
\eeq
where $\eta_{\mu\nu}=$ diag(-1,1,1,1) and
the coordinates parallel to the brane are denoted by $x^{\mu}=(t,x^i)$,
while $y$ is the coordinate transverse to the brane.

\vspace{.3cm}
For the bulk action,
%$$S=S_{\rm g}+S_{\rm b},$$
the BPS solution is given by solving the first order equations \cite{ST,DFGK},
\beq
 \phi_I'={v\over 2}{\partial W\over\partial\phi_I}, \qquad
    {A'\over A}=-{v\over 3}W,  \label{first-order}
\eeq
where $'=d/dy$. It is known that these equations are
the necessary conditions for the supersymmetry of the
solution. And the solutions of
(\ref{first-order}) satisfy the equations of motion for any $W$.

Also at the position of the brane, the equations
(\ref{first-order}) are satisfied due to the special form of the
brane action as given in (\ref{baction}), which is taken such that
we can preserve the form of the supersymmetric or BPS bulk
solutions \cite{BKP,BD}.

\vspace{.3cm} Our purpose is to find an effective potential
$V^{\rm eff}(\phi)$ for scalars in a simple bulk background to
obtain a solution, instead of solving the above equation
(\ref{first-order}). The metric is restricted to be of the AdS$_5$
form, which corresponds to $A=e^{-|y|}$.
% by setting the parameters as $\kappa^2=2$ and $v=-2$.
%%%%%%%%%%%%% Modified 2 June
%The strategy to obtain the effective potential
%is as follows. First, 
It will be possible to estimate the effective brane action 
$S^{\rm eff}_b$
from the viewpoint of AdS/CFT correspondence as in the way
\cite{Gidd,GKa,HeSk,GY}, but it would be very complicated due to
the estimation of the bulk part. So, here, 
we consider a situation where it would be possible to
estimate $S^{\rm eff}_b$ from 
gravity side classical potential as follows.
%\beq
% S^{\rm eff}_b={1\over 2}S_{\rm b} + \ln Z_5(g)
%           %= {1\over 2}\tilde{S}_b + S_{\rm CT}+S_{\rm CFT}
%  \label{effaction-1}
%\eeq
%\beq
%  Z_5(g,\chi)=\int DG D\phi e^{iS_{\rm g}},   \label{effective-1}
%\eeq where $S_{\rm b}$ and $S_{\rm g}$ are given above. 
We
pull the hidden brane from $y=y_h$ to the position of the visible
brane, $y=y_v$, by the brane running method. As a result the bulk 
space vanishes, then we obtain 
\beq
 S^{\rm eff}_b={1\over 2}\tilde{S}_{\rm b} \, ,  \label{effaction3}
\eeq 
where $\tilde{S}_{\rm b}$ denotes the sum of the actions of
hidden and visible branes after the running, as mentioned above.
Then we can estimate the
effective potential to be \beq
 V^{\rm eff}=-{1\over 2}\tilde{S}_{\rm b} \, ,  \label{effaction4}
\eeq and the solution of $\phi$ must be studied in order to estimate
the above effective action.
%%%%%%%%%%%%%%%%%%%%%%%%%%%%%%%%

\section{Brane running and BPS}

Consider the case of one scalar $\phi$. Then the equation for it is
obtained as \beq
  {\phi}''+4{A'\over A}{\phi}'
           +{q^2\over A^2}\phi={\partial V(\phi)\over \partial \phi}
                -2{\partial W(\phi)\over \partial \phi}\delta(y),
                         \label{chi-eq}
\eeq where $q^2$ is the four dimensional momentum square of $\phi$
and we take $y_h=0$ for simplicity. The boundary condition for
$\phi$ is written as 
\beq
  \phi'(0)=-{\partial W(\phi)\over \partial \phi}|_{y=0}. \label{bound-1}
\eeq We extend this equation to the position $y>0$  where the
running brane arrived, by introducing the action of this running
brane as \beq
    S_{\rm b}^{\rm (R)} = -2\int d^4x \sqrt{-g}
          \sum_{n=0}{\phi^n\over n!}\tau^{(n)}(y), \label{baction-R}
\eeq
where $0<y<y_v$. 
%%%%%%%%%%%%%%%%%%%%%%%%%%%% Modified 2
We should notice the following point for this brane action.
In general, many kinds of derivative terms, kinetic and higher derivative
terms, appear in the effective action 
after a running, but here only the non-derivative terms are retained in order
to find a vacuum state which should be dominated by low frequency modes.

\vspace{.3cm}
The coefficients for each power of the scalar field $\phi^n$ in 
(\ref{baction-R}) are
defined as the running coupling constants, $\tau^{(n)}(y)$, since they 
in general vary with $y$. This variation is determined by the bulk
configuration of the fields which are the solution of the classical
equations including (\ref{chi-eq}). On the other hand, the running behavior
of $\tau^{(n)}(y)$ is considered as a renormalization group equation
obtained due to 
the interactions with the dual field theory of the bulk background. 
In this sense, the approach considered here
provides a quantum information for the brane
action in a sense of gauge/gravity correspondense. Then
it would be natural to consider the brane action
obtained after a brane running as an effective action in which the
quantum effects of the dual theory are reflected through $\tau^{(n)}(y)$. 
So we can see the dynamical properties of the dual gauge theory
from the $y$-dependence of $\tau^{(n)}(y)$.

The interesting point of this brane running method is that
they are determined by the boundary condition defined at the arrived new
position, $y_h=y$, of the hidden brane.
It is given by using (\ref{baction-R}) as
%%%%%%%%%%%%%%%%%%%%%%%%%
\beq
  \phi'(y)=\sum_{n=1}{\phi^{n-1}\over (n-1)!}\tau^{(n)}(y) . \label{boundchiy}
\eeq
%%%%%%%%%%%%%%%%%%%%%%%%%%% Modified 3
Since the boundary condition at $y=0$ is written by the potential
$W(\phi)$, which is used at the initial point $y=0$,
then the initial values, $\tau^{(n)}(0)$, are determined by $W$.
Expanding $W$ as
\beq
   W(\phi)=\sum_{n=0}{\phi^n\over n!}W^{(n)}, \qquad 
        W^{(n)}={d^n W\over d\phi^n},
\eeq
and considering (\ref{boundchiy}) at $y=0$, we obtain
the boundary condition for $\tau^{(n)}(y)$ as
\beq
  \tau^{(n)}(0)= -W^{(n)}.
        \label{initial-condition}
\eeq
Further, for the region of $0<y<y_v$, we obtain $\tau^{(n)}(y)=-W^{(n)}$
from (\ref{first-order}) and (\ref{boundchiy}), then
\beq
  \tau^{(n)}(y)=\tau^{(n)}(0).
        \label{initial-condition2}
\eeq

\vspace{.3cm}
Therefor, for any BPS solution the coefficients $\tau^{(n)}(y)$ 
are at the fixed point. Then, the potential of hidden part coincides
with the one of visible brane with opposite sign
when the hidden brane arrived at the position of the visible brane. 
After all, we arrive at the following result
\beq
 V^{\rm eff}=0 \, .  \label{effaction5}
\eeq
Then, the potential is independent of the interbrane distance 
and it is arbitrary in this case. In other words,
it can not be determined.
In order
to obtain some non-zero effective potential, we should consider a 
non-BPS background.
%%%%%%%%%%%%%%%%%%%%%%%%%%%%%%%%%%%%%%%%%%%%%%%%%

\vspace{.4cm} 
\section{Non BPS case}
Here we derive a non-BPS solution by 
solving directly the second order equation (\ref{chi-eq}) instead of
the first order equations
(\ref{first-order}). Then, the BPS condition is 
broken. Namely, we fix the bulk as AdS$_5$: \beq A'/A=-1 \; ,
\quad \quad \tau^{(0)}(y)=-W^{(0)}(0)=-{3\over2} \; . \label{AdS5}
\eeq Further we assume \beq
  \tau^{(1)}(y)=0
\eeq
to make it possible to solve the flow equation.
Other $\tau^{(n)}(y)$ are obtained from the brane running method.

%%%%%%% added
In Ref.\cite{LMS}, tadpole and quadratic terms are considered in a
truncated form by introducing a fixed point of the brane running.
In this sense, we are considering a different model.
%%%%%%%%%
The trivial BPS solution, $\phi=0$, is consistent with these
conditions, but we search for a non-trivial solution. The
consistency of such a non-trivial solution and the background
AdS$_5$ would be justified when the back-reaction is negligibly
small, that is, for sufficiently small $\phi$. Thus, this
derivation is based on the perturbation.

%%%%%% Modified %%%%%%
The differential equations for $\tau^{(n)}$ for $n\geq 2$ are
obtained as follows. Differentiate (\ref{boundchiy}) with
respect to $y$, and rewrite $\phi''$ and $\phi'$ in terms of $\phi$
by using (\ref{boundchiy}) and (\ref{chi-eq}). And,
expand $W(\phi)$ and $V(\phi)$ in the series of $\phi$ as
$%\beq
 W(\phi)=\Sigma_n{1\over n!}W^{(n)}\phi^n,~
 V(\phi)=\Sigma_n{1\over n!}V^{(n)}\phi^n
$%\eeq
, where %$W^{(i)}\equiv {\partial^i W\over \partial \phi^i}$ and
$V^{(n)}\equiv {\partial^n V\over \partial \phi^n}$. Then, 
by observing the coefficient of $\phi^n$, we obtain the following
results,
%%%%%% Modified %%%%%%
\beq
  (\tau^{(n)})'=-4{2\over 3}W^{(0)}{(0)}\tau^{(n)}+V^{(n)}
 -\sum_{m=2}^{n}(n-1)!{\tau^{(m)}\tau^{(n+2-m)}\over (m-2)!(n-m+1)!}
    +{q^2\over A^2}\delta_{n2}, \label{diff-eq}
\eeq
\beq
   V^{(n)}\equiv {\partial^n V\over \partial \phi^n}=
     -4{2\over 3}W^{(0)}{(0)}W^{(n)}
   +\sum_{m=2}^{n}n!{W^{(m)}W^{(n+2-m)}\over 2(m-1)!(n-m+1)!}
-{4\over 3}\sum_{m=2}^{n-2, n\geq 4}n!{W^{(m)}W^{(n-m)}\over m!(n-m)!}
\eeq
for $n\geq 2$.

\vspace{.4cm} Here we consider  the small $q^2$ region to see the
effective potential, so we neglect the term ${q^2\over A^2}$ in
(\ref{diff-eq}). In this case, we find \beq
  \tau^{(2)}(0)'=\tau^{(3)}(0)'=0, \quad \tau^{(4)}(0)' 
     =-8(\tau^{(2)}(0))^2 .
\eeq These results imply that $\tau^{(2)}$ and $\tau^{(3)}$ are
not running, but $\tau^{(4)}$ is running and should decrease with
$y$ for any $W$. This means that the effective potential behaves
as $V^{\rm eff}\propto -\phi^4$, so the braneworld is unstable for
the present model. This instability is related to the breaking of
the  BPS conditions. In the next section, we shall see the details
of this point in terms of a simple model of gauged supergravity.

\subsection{Modified superpotential model}

Here we consider a truncated superpotential \cite{FGPW,ST}
\beq
 W=-e^{-2\phi/\sqrt{6}}\left(1+{1\over 2}e^{6\phi/\sqrt{6}}\right) .
\eeq
For this potential, $\tau^{(i)}$ for $i\leq 3$ are not
running as shown above and we find
\beq
  \tau^{(2)}=2, \quad \tau^{(3)}={4\over \sqrt{6}}.
\eeq
For $i\geq 4$, $\tau^{(i)}$ are running and we obtain the next two
terms as follows
\beq
  \tau^{(4)}=8(e^{-4y}-1/2), \quad
    \tau^{(5)}={40\over 9\sqrt{6}}e^{-6y}
      \left(32-36e^{2y}+7e^{6y}\right).
\eeq
Up to this order $V^{\rm eff}$ is written as
\beq
   V^{\rm eff}=A^4\left({\Delta\tau^{(4)}\over 4!}\phi^4+
       {\Delta\tau^{(5)}\over 5!}\phi^5\right),
\label{W} \eeq
where we set $y_v=y$ and $\Delta\tau^{(i)}\equiv
\tau^{(i)}(y)-\tau^{(i)}(0)$. This potential has a minimum at
$$\phi=-4\Delta\tau^{(4)}/\Delta\tau^{(5)}\equiv <\phi>,$$
which is negative, and we see that $|<\phi>|\geq 3\sqrt{6}/5 > 1$.
Then this minimum point is outside the perturbation regime. In
this sense, we can say only that the braneworld solution given
here is unstable since the potential $V^{\rm eff}$ indicates the
existence of other stable points.

\vspace{.3cm} Another point to be noticed is the $y$-behavior of
$V^{\rm eff}(<\phi>)$. When it shows a non-trivial minimum at some
finite $y$, it indicates the stabilization of the distance between
the two branes. Although the value of $<\phi>$ given above is
outside  the approximation region considered here, we could study
the $y$-dependence of $V^{\rm eff}(<\phi>)$ for this case.
%%%%%%%%%%%%%%% Fig %%%%%%%%%%%%%%
\begin{figure}[htbp]
\begin{center}
\voffset=15cm
  \includegraphics[width=9cm,height=7cm]{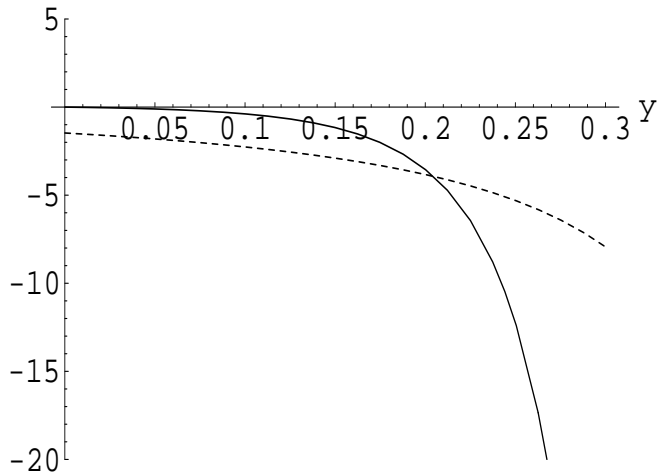}
\caption{The solid curve represents the potential, $V^{\rm eff}(<\phi>)$, for the BPS brane solution up to $\phi^5$.
The dashed curve represents $<\phi>$.
\label{mxigraph1}}
\end{center}
\end{figure}
%%%%%%%%%%%%%%% Fig %%%%%%%%%%%%%%
As shown in  Fig.1, it indicates no stable point. Then, we find
again that this solution is unfavorable.

\vspace{.3cm}
 Next, we consider a possible mechanism to stabilize
the braneworld. It is realized by adding a small mass term,
$-\epsilon \phi^2/2$, to the brane action:
 \beq
    S_{\rm b} = -{v}\int d^4x dy\sqrt{-g}\left\{W(\phi)
     -{1\over 2}\epsilon\phi^2\right\}
\left(\delta(y-y_h)-\delta(y-y_v)\right). \label{baction-mod}
\eeq
In this case, only the boundary condition for $\tau^{(2)}$ is
modified as $\tau^{(2)}(0)=-(W^{(2)}(0)-\epsilon)$. Then $\tau^{(2)}$
is running and we obtain
\beq
 \tau^{(2)}=2+{\epsilon\over 1+\epsilon y},
\eeq
$$ %\beq
 \tau^{(3)}={1\over \sqrt{6}(1+\epsilon y)^3}
 \left\{ 3\epsilon e^{-2y}(2-2\epsilon+\epsilon^2)
+4+6(-1+2y)\epsilon+6(1-2y+2y^2)\epsilon^2\right. \nonumber
$$ %\eeq
\beq
   \left.~~~~~~~~~~~~~~~~~~~~
      +(-3+6y-6y^2+4y^3)\epsilon^3\right\},
\eeq up to  order  $\phi^3$. To this order, the effective
potential is written as \beq
   V^{\rm eff}=A^4\left({\Delta\tau^{(2)}\over 2}\phi^2+
       {\Delta\tau^{(3)}\over 3!}\phi^3\right), \label{potential-mod}
\eeq
and it has a minimum at $<\phi>=-2\Delta\tau^{(2)}/\Delta\tau^{(3)}$.
For small $\epsilon$, we have
\beq
   <\phi>\sim -\sqrt{{2\over 3}}~ {y\over 1-e^{-2y}} \epsilon .
\eeq
Hence, a small value of $<\phi>$ can be realized,
as long as $\epsilon$ is small enough and $y$ is not large.

\vspace{.3cm} The next problem is to see the $y$-dependence of
$V^{\rm eff}(<\phi>)$. The result is shown in  Fig.2.
%%%%%%%%%%%%%%% Fig %%%%%%%%%%%%%%
\begin{figure}[htbp]
\begin{center}
\voffset=15cm
  \includegraphics[width=9cm,height=7cm]{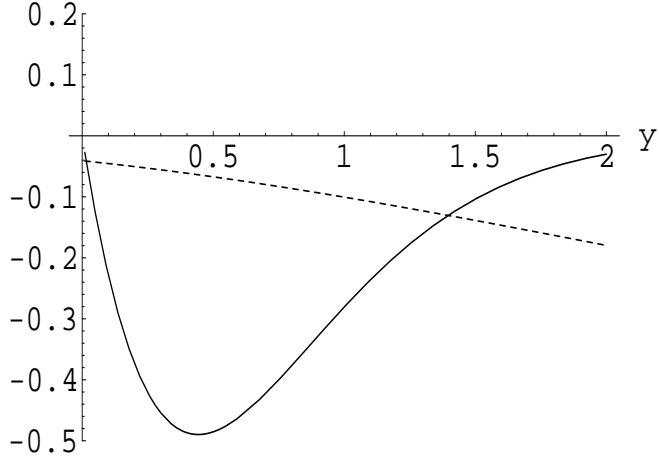}
 \caption{The solid curve represents the potential, $V^{\rm eff}(<\phi>)\times
10^{6}$, for the non BPS brane solution up to $\phi^3$.
The dashed curve represents $<\phi>$. Here $\epsilon =0.1$.
\label{potengraph1}}
\end{center}
\end{figure}
%%%%%%%%%%%%%%% Fig 2 %%%%%%%%%%%%%%
As seen from this figure,  $V^{\rm eff}(<\phi>)$ has a minimum at
$y \sim 0.4$ where $<\phi>\sim 0.06 <<1$ as expected. The figure
is shown for $\epsilon=0.1$, but this behavior  changes only
little with varying $\epsilon$ and its sign is not important. Then
the small modification of the brane action given here is enough to
stabilize the braneworld solution with an interbrane distance of
$y \sim 0.4$. The back-reaction from this modification will be
negligible since we can take very small $\epsilon$.

%\vspace{.3cm}
\subsection{Goldberger-Wise model} 
Let us briefly consider
the Goldberger-Wise model \cite{GW} and its stabilization via the
brane running. 
%%%%%%%%%%%%%%%%%%%%%%%%
The brane running method has already been applied to this 
model with success in reproducing the stabilization~\cite{LMS},  
indicating that 
the brane running method is a reasonable way of evaluating the effective 
action. In the previous analysis, only the brane running 
(renormalization group flow) near 
the fixed point is considered. Our reanalysis has no such restriction. 
%%%%%%%%%%%%%%%%%%%%%%%%%

The brane action of the model is \beq
    S_{\rm b} = -{v}\int d^4x dy\sqrt{-g}
\left(W_{h}\delta(y-y_h)-W_{v}\delta(y-y_v)\right) \; ,
\label{GW-brane} \eeq where
$W_{x}(\phi)=(\phi^2-\alpha_{x})^2+\beta_{x}$ and
$W_{x}(0)=\alpha_{x}^2+\beta_{x}=-3/2$ for $x=h$ and $v$, and the
bulk potential $V$ is defined as $V=-6+m^2\phi^2/2$. This model
obviously breaks the bulk supersymmetry and the BPS conditions,
since $V$ and $W_{x}$ defined above do not satisfy
(\ref{superPot}). It is found that $W_{h}^{(2)}=-4\alpha_{h}$,
$W_{v}^{(2)}=-4\alpha_{v}$, $W_{h}^{(4)}=W_{v}^{(4)}=4!$ and
$V^{(2)}=m^2$. So $\tau^{(2)}(y)$ is obtained as
\beq \tau^{(2)}(y)={
\lambda_{1}(\lambda_{2}+W_{h}^{(2)})e^{\lambda_{1}y}-
\lambda_{2}(\lambda_{1}+W_{h}^{(2)})e^{\lambda_{2}y} \over
(\lambda_{2}+W_{h}^{(2)})e^{\lambda_{1}y}-
(\lambda_{1}+W_{h}^{(2)})e^{\lambda_{2}y} } \; \eeq where
$\lambda_{1}=2+\sqrt{4+V^{(2)}}$,
$\lambda_{2}=2-\sqrt{4+V^{(2)}}$. Furthermore, it is found from
the initial conditions $W_{h}^{(3)}=0$ and $W_h^{(4)}=4!$ that
$\tau^{(3)}(y_v)=0$ and $\tau^{(4)}(y_v) \neq 0$, that is, \beq
\tau^{(4)}(y_v)=-W_{h}^{(4)} \exp \left\{
\int_{y_h}^{y_v}4(1-\tau^{(2)}) dy \right\} \; . \eeq The explicit
form of $\tau^{(4)}(y_v)$ is lengthy, so it is not shown here.
Assuming that $\phi$ is small, we then consider the brane solution
up to $\phi^4$. Up to this order, the effective potential is \beq
   V^{\rm eff}=A^4\left({\Delta\tau^{(2)}\over 2!}\phi^2+
       {\Delta\tau^{(4)}\over 4!}\phi^4\right),
\label{GW-V} \eeq where
$\Delta\tau^{(i)}=\tau^{(i)}(y_v)+W_{v}^{(i)}$ for $i=2,4$, and it
has a minimum at $<\phi>^2=-3!\Delta\tau^{(2)}/\Delta\tau^{(4)}$.
When parameters are taken as $\alpha_{h}=O(m^2)$ and
$\alpha_{v}=O(m^2)$, it is easily found that
$\Delta\tau^{(2)}=O(m^2)$ and $\Delta\tau^{(4)}=O(1)$, leading to
$<\phi>^2=O(m^2)$ and $V^{\rm eff}(<\phi>)=O(m^4)$. Hence, the
back-reaction is negligible for small $m^2$.
%%%%%%%%%%%%%%% Fig 3 %%%%%%%%%%%%%%
\begin{figure}[htbp]
\begin{center}
\voffset=15cm
  \includegraphics[width=8cm,height=6cm]{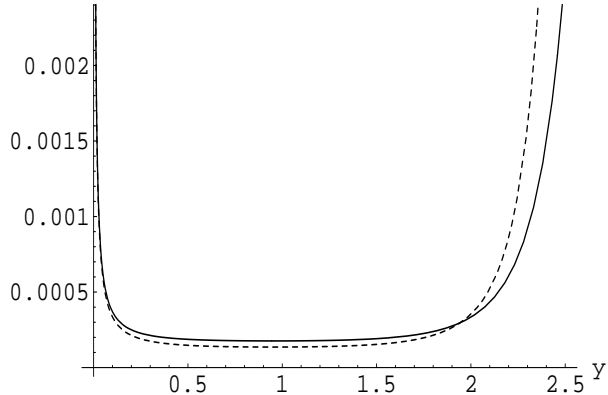}
\caption{The solid curve represents $V^{\rm eff}(<\phi>)\times
10^{4}$ for the brane solution up to $\phi^4$.
The dashed curve represents $<\phi>^2$. Parameters taken here are
$\alpha_{h}=5m^2/4$, $\alpha_{v}=m^2/4$ and $m^2=10^{-4}$.
\label{potengraph2}}
\end{center}
\end{figure}
%%%%%%%%%%%%%%% Fig 3 %%%%%%%%%%%%%%
Figure 3 shows $V^{\rm eff}(<\phi>)$ and $<\phi>^2$
for the case of
$\alpha_{h}=5m^2/4$, $\alpha_{v}=m^2/4$ and $m^2=10^{-4}$.
The effective potential has a minimum at $y \sim 1.0$
where $<\phi>^2\sim m^2 \ll 1$.
The effective potential thus obtained would
include some sort of quantum corrections
in the sense that $\tau^{(2)}$ and $\tau^{(4)}$
are running in the renormalization group procedure.
As an interesting point, the effective potential shows
a concave shape which is also seen in the classical limit
of $V^{\rm eff}(<\phi>)$ as reported in the original work \cite{GW} of this
model.

\vspace{.3cm}
Lastly, we make a comment on another possible origin to
stabilize the interbrane
distance ~\cite{brevik01,brevik02}.
%the alternative to the
%Goldberger-Wise stabilization method
%mentioned above. There exists an alternative to the method, namely to
%introduce
Consider the temperature on the brane and make use of
the finite temperature field theory for this system.
In this case, the Casimir energy between two branes
for fields in the bulk contributes to the
free energy, and its fermionic part could give a minimum of the free energy
at a finite interbrane distance. Then,
%When $TL \ll 1$ for the brane temperature $T$ and the AdS radius $L$,
the effective potential could have a nontrivial minimum even for
$<\phi>=0$ contrary to the case of the Goldberger-Wise
stabilization method mentioned above. However, we notice that
both stabilization mechanisms would be compatible.
%The order of magnitude of the warp factor $a=\exp{(-y/L)}$ at the minimum,
%$a=a_m$, is given roughly by $ a_m \sim {T L} \ln ({1/TL})$.
%Taking $y/L \simeq 12$ means that the visible brane is
%associated with the TeV region, even if $L$ is the Planck scale.

%\newpage
%\vspace{2cm}
%%%%%%%%%
%%%%%%%%%%%%%%%%  Conclusion %%%%%%%%%%%%%%%%%%
\section{Concluding remarks}

We study the interbrane distance of the two brane system with a
bulk scalar field. It is examined
in terms of the effective potential obtained by
the method of brane running. Here, the bulk scalar is
considered in the gauged supergravity and it couples to
the brane in a form consistent with the BPS conditions for the
bulk solutions. In this
%%%%%%%%%%%%%%%%%%%%%%%%%%%%%%%% Modified 5
case, for bulk BPS solutions, the parameters in the potential 
of the hidden brane are not running even if the brane position is changed
by the brane running. Then the effective
potential, which is defined as the sum of two branes,
becomes zero when the hidden branes arrived at the position of the
visible one. As a result, the effective potential
%%%%%%%%%%%%%%%%%%%%%%%%%%%%%%%%%%%%%%%%%
is independent of the value of the interbrane distance. Thus, the
interbrane distance is arbitrary in this case. This property is
true for any type of superpotential and BPS solutions. So it would
be necessary to consider a non-BPS solution to obtain a finite and
stable interbrane distance. We can show such two non-BPS examples.

\vspace{.3cm}
We consider the case that one of the BPS conditions is
broken by considering a nontrivial solution for the scalar $\phi$,
while the bulk configuration keeps AdS$_5$ intact. This solution
is justified when $\phi$ is small. In this case, we find that the
effective potential behaves as $V^{\rm
eff}=-8(W^{(2)})^2\phi^4+O(\phi^5)$, and there is no stable
point for the interbrane distance in this effective potential
within the  small $\phi$ approximation.
%%%%%%%%%%%%%%%%%%%%%%%%% Modified 6
We find that
%%%%%%%%%%%%%%%%%%%%%
the braneworld of a non-trivial scalar can be stabilized for small
$\phi$ when a small BPS breaking term is added to the brane
action. We could show this by choosing the superpotential given in
\cite{FGPW,ST} as an example. The stable interbrane distance
obtained in this way is not large. However, if necessary, it may
be possible to get a larger stable distance by breaking the BPS
conditions more severely. Actually, when we break the BPS
conditions for both brane and bulk actions as in the Goldberger -
Wise approach \cite{GW}, we succeed as the second example
in making a stable braneworld with a larger interbrane distance.

%%%%%%%%%%%%%%%%%%%%%%%%%%%%%%%%
\section*{Acknowledgments}
This work has been supported in part by the Grants-in-Aid for
Scientific Research (13135223, 14540271)
of the Ministry of Education, Science, Sports, and Culture of Japan.

\end{document}